**Emergence and relaxation of an e-h quantum liquid phase in photoexcited MoS₂ nanoparticles at room temperature**

*Pritha Dey, Tejendra Dixit, Vikash Mishra, Anubhab Sahoo, Cheriyanath Vijayan and Sivarama Krishnan \**

Pritha Dey, Anubhab Sahoo, Prof. Cheriyanath Vijayan, Prof. Sivarama Krishnan

Department of Physics, Indian Institute of Technology Madras, Chennai 600036, India

Prof. Sivarama Krishnan

Quantum Center for Diamond and Emerging Materials, Indian Institute of Technology Madras, Chennai 600036, India

Dr. Tejendra Dixit

Optoelectronics and Quantum Devices Group, Department of Electronics and Communication Engineering, Indian Institute of Information Technology Design and Manufacturing Kancheepuram, Chennai, 600127, India

Dr. Vikash Mishra

Department of Physics, Nano Functional Materials Technology Center and Materials Science Research Center, Indian Institute of Technology Madras, Chennai 600036, India

ABSTRACT

Low-dimensional transition metal dichalcogenide (TMDC) materials are heralding a new era in optoelectronics and valleytronics owing to their unique properties. Photo-induced dynamics in these systems has mostly been studied from the perspective of individual quasi-particles – excitons, bi-excitons or, even, trions - their formation, evolution and decay. The role of multi-body and exciton dynamics, the associated collective behaviour, condensation and inter-excitonic interactions remain intriguing and seek attention, especially in room-temperature scenarios which are relevant for device applications. In this work we evidence the formation and decay of an unexpected electron-hole quantum liquid phase at room-temperature on ultrafast picosecond timescales in multi-layer $MoS_2$ nanoparticles through femtosecond broadband transient absorption spectroscopy. Our studies reveal the complete dynamical picture: the initial electron-hole plasma (EHP) condenses into an quantum electron-hole liquid (EHL) phase which typically lasts as long as 10 ps, revealing its robustness, whereafter the system decays through phonons. The formation of the EHL phase has an important consequence: the strong interaction between the excitons decreases Coulomb screening leading to a renormalization of single exciton energies. Our measurements capture this essential bandgap renormalization (BGR) in transient absorption spectra over a wide range of pump fluences; the BGR can be as high as 50 meV, as discerned from the time-dependent shift in excitonic resonances. Although the total BGR is a cumulative effect of the quantum EHL phase along with the EHP phase and phonons, we employ a successful physical model to extract each of these contributions using a set of coupled nonlinear rate equations governing the individual population of these constituent phases. Beyond the observation of the electron-hole liquid-like states at room-temperature, this study reveals a new generic feature, the EHL phase, in the ultrafast dynamics of photo-excited low-dimensional systems arising out of the collective many-particle behavior and correlations.

## 1. Introduction

The rich dynamics of light-matter interaction in semiconducting systems goes beyond mere exciton formation when the excitation density is raised – stronger excitations lead to excitonic quasi-particle complexes such as bi-excitons[1,2] and trions[3]. Under suitable conditions the condensation of these excitonic particles into quantum clusters such as electron-hole liquid (EHL) phase can occur[4,5]. The distinguishing behavior of the quantum nature of EHL in contrast to typical a "classical" liquid is explored in coupled semiconductor quantum well recently[6]. These quantum phases play a crucial role in realizing room temperature applications exploiting the unique properties of low-dimensional transition metal dichalcogenides (TMDCs) such as $MoS_2$ and realizing devices out of them[7,8]. Therefore, insights into the dynamics of these condensates on the natural ultrafast timescales of their evolution are the key to such device realizations. The formation and recombination pathways of single excitons have been explored in a number of studies[9,10], however the condensed excitonic phases such as EHL, especially at room-temperature, remain scarcely accessible due to the low critical-temperature essential for their formation in bulk or even most nanoscale systems. Owing their high binding energies, recent reports have evidenced the existence of stable EHL phases in low-dimensional TMDCs -monolayers[11,12] and heterostructures[13]. This has opened up new challenges and opportunities towards applying and exploiting these systems in various scenarios.

Since their prediction by Keldysh in 1968[14], the quantum EHL phase has been realized in several semiconductor systems[5,15–22]; large exciton concentrations with inter-excitonic distances approaching the excitonic Bohr radius are necessary for these observations. Spectroscopy of conventional semiconductors[5,15–22], quantum wells[23,24] and superlattices[25] has revealed corresponding shifts in exchange and correlation energies leading to a phase transition from exciton gas to the EHL, especially at low-temperatures (~ few Kelvins). Hence, the EHL phase, unlike classical macroscopic states, is unstable at room temperature in most semiconducting[26] systems. Consequently, in such systems the unique aspects of the EHL phase such as metallic conductivity remains limited in applicability for possible devices. However, low-dimensional TMDCs open new doors for the realization of the EHL phase at room temperature owing to the unique properties they combine - high exciton binding energy ~ 0.5 eV, low dielectric constant $\varepsilon$" ~ 8, large carrier effective masses $m^* \sim 0.5 \, m_e$ [27] and longer exciton lifetimes. Although earlier studies have explored the steady-state properties of such condensates e.g. the phase diagram[12], size of EHL droplets etc.[28].; the transient properties of this phase, which are necessary to tailor the device performances, remains quite unexplored so far.

In this work, we evidence the formation of an EHL phase on picosecond timescales in the ultrafast relaxation of photoexcited multilayered $MoS_2$ nanoparticles. This is revealed by femtosecond transient absorption spectroscopy (fs-TAS). This enabled us to measure bandgap renormalization (BGR) with fs time resolution which revealed the existence of this condensed phase. This is the first time, to the best of our knowledge, time-resolved dynamics of BGR arising from the highly correlated EHL phase has been measured and revealed. Although the theoretical estimation of BGR due to the EHL state was reported earlier[25], experimental evidence for EHL-induced BGR has not been reported so far, to the best of our knowledge. Through fluence dependent excitonic resonance shifts evident in fs-TAS, we evidence this feature and follow it from formation to decay in the ultrafast relaxation of this multilayer TMDC system. These are good hosts for observing condensed excitonic phases due to their relatively long excitonic lifetimes. The interlayer interactions in the multilayered $MoS_2$ results in a pronounced indirect nature of the band structure[29,30] leading to longer exciton lifetimes compared to their monolayer direct bandgap counterpart[11,31,32]. Furthermore, we use a facile technique to prepare multilayer $MoS_2$ NPs and held them suspended in water, as detailed in the Methods section of this article. These are an excellent choice for studying liquid-like excitonic many-body complexes[33]. We induce the dynamics, with pump pulses having a fluence of a few mJ·cm$^{-2}$ initiating the photoexcitation and formation of a dense e-h plasma in this system. As the system relaxed on ~ps timescales, we investigated the novel multi-particle excitonic phases in these nanoscale TMDCs in suspension. Bandgap renormalization effects in the aggregated condensed phase are hitherto largely unexplored in such systems. By incorporating ultrafast transient absorbance measurements, high-quality data acquisition along with robust analysis and an intuitive physical model for the ensuing dynamics, we evidence unambiguous signatures of the EHL phase in the dynamics of relaxation, carrier population and bandgap renormalization. The remainder of this article discusses this in detail.

We use a wide range of pump fluences spanning over 40 times the lowest applied pump fluence in magnitude, up to several mJ·cm$^{-2}$, to investigate time-resolved changes in the renormalization and carrier dynamics due to the formation of different excitonic aggregates and condensates as the system relaxes. According to previous studies[37], an EHP state is produced following primary photoexcitation as a result of the screening of Coulomb attraction between electrons and holes at high carrier densities. Since screening reduces the binding energy of excitons they remain dissociated[16,38,39]. This results in a bandgap renormalization (BGR) which is captured in the time-resolved spectra recorded by the fs-TAS, manifesting as a blue-shift in the excitonic energy characteristics[40]. BGR however affects the whole band structure which manifests in an overall shift which involves both excitonic resonances and band-nesting regions. We observe that the restoration of the bandgap to its original band structure is gradual at higher pump fluences. To understand the dynamics of each of the excitonic phases including EHP, EHL and exciton gas along with phonons, we analyze the bleach signal corresponding to the B exciton, ~ 610 nm, on the basis of two parameters in the absorption spectrum as a function of pump-probe time delay, namely, the amplitude of the negative going excitonic peak ($\Delta A$) and the shift in its position ($\Delta E$). While the amplitude correlates with the instantaneous total population of the excitonic species of interest; the renormalized bandgap is reflected in the energy shift of the bleach features. We acquire several datasets of transient absorption $\Delta A$ as a function of pump-probe time delay, $\Delta t$, for a series of pump fluences $F$, from $F$ = 0.26 mJ·cm⁻² , corresponding to e–h pair density of $9.8 \times 10^{18}$ cm⁻³ to $F$ = 10.2 mJ·cm⁻² where this density rises to $5.24 \times 10^{19}$ cm⁻³ (cf. the Supplement S2 for details of carrier density calculations). Given the pump fluences and the corresponding e-h densities, we note that carrier concentration is varied from well below to far *above* the Mott-density for multilayered MoS₂[27] (which is ~ $6 \times 10^{12}$ cm⁻² for the 2D layers and estimates up to ~$1.47 \times 10^{19}$ cm⁻³ in the 3D limit).

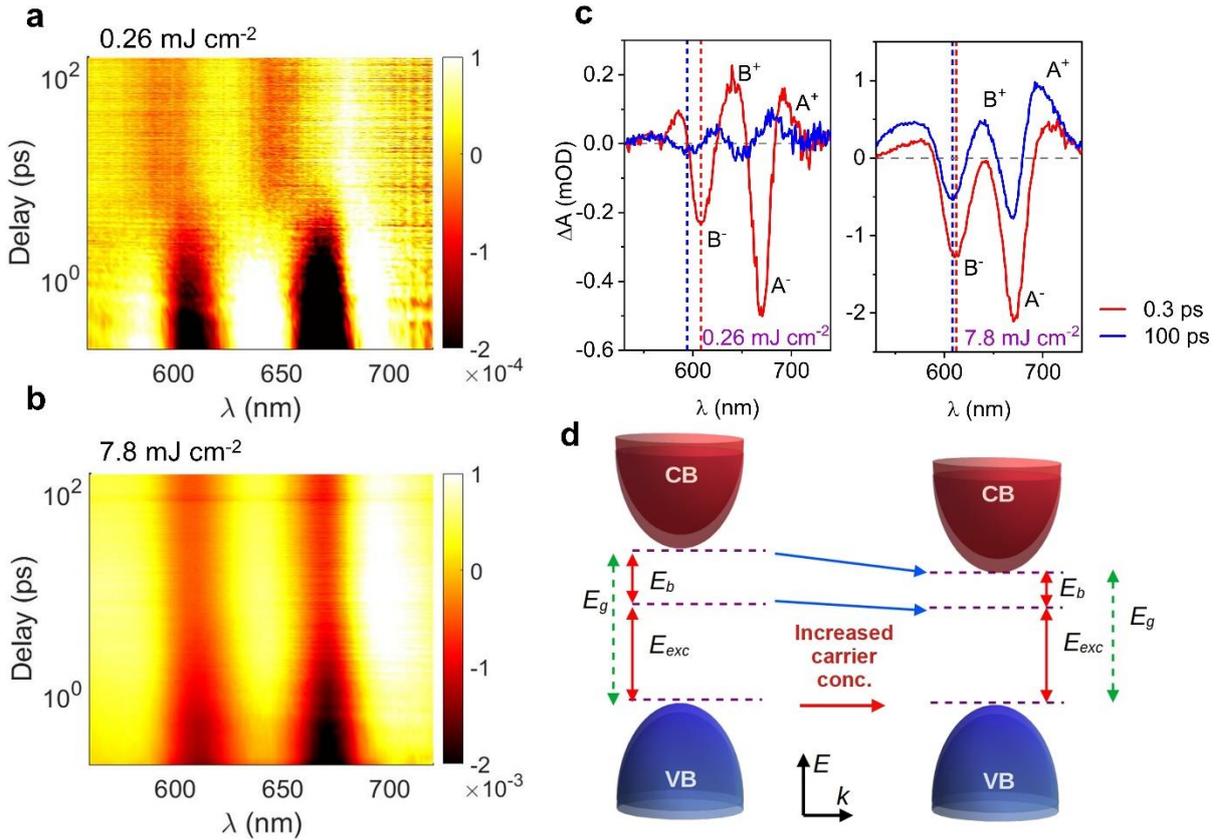

**Figure 1.** (( **Transient absorption map and the time dependent renormalization of band gap: a**, 2D pseudo-color spectral map of transient absorption spectra (TAS), the horizontal axis

is the wavelength and the vertical axis the delay between the pump and probe pulses while the differential absorption of the probe is represented using a color scale as shown in the accompanying bar. We measured this for a pump fluence of 0.26 mJ·cm$^{-2}$. This evidently shows a drastic shift in the maximum of the bleach feature for a probe delay of about 4 ps. This shift arises from the abrupt change in the excitonic resonance during the relaxation of the multilayer MoS$_2$ system on ultrafast timescales. **b**, the corresponding 2D spectral map from fs-TAS at high-fluence of 7.8 mJ·cm$^{-2}$ revealing almost negligible change in the peak position of the bleach feature, compared in the panel **a**. This is indicative of very little change in excitonic resonance with delay in this case, as the system relaxes. **c**, TAS spectra at two pump fluences 0.26 mJ·cm$^{-2}$ and 7.8 mJ·cm$^{-2}$ at two delay times of 0.3 ps (red) and 100 ps (blue), respectively. This clearly evidences shift in the excitonic bleach minima, the bleach feature, at these delays marked using the red- and blue-dashed lines, respectively. **d**, Illustration of the bandgap renormalization (BGR) mechanism discussed in the text, exhibiting the reduction in bandgap and the corresponding exciton binding energy with increasing carrier density.))

We excited these multilayer MoS$_2$ NPs first using 500 nm (2.48 eV) 100 fs pump pulses, well above the A and B excitons of the conventional MoS$_2$ system, and recorded spectra transient absorption spectra at various time delays between the pump and the probe pulses to realize time-resolved femtosecond transient absorption spectra[36], fs-TAS, cf. the "Methods" section and section 3 of the Supplement for details. Typical transient absorption spectra exhibit two negative going bleach features centered at 610 nm and 670 nm which can be attributed to bleach signals corresponding to the B and A excitons, respectively; these features arise from the increased occupation of electrons in the conduction band and holes in the valance band leading to a reduction in probe absorption at the K-point. Figure **1a**. presents the two-dimensional transient absorption map recorded with a pump fluence of 0.26 mJ·cm$^{-2}$ where a drastic shift in the bleach energy corresponding to excitons A and B is evident at a delay of ~ 4 ps indicating the transition from EHPs to excitons thereby more or less relaxing the band structure to its original state. In contrast, at a much higher pump fluence of 7.8 mJ·cm⁻², panel **1b**, the corresponding peaks in the identical bleach feature show almost negligible change in their position evidencing a slower recovery of the band structure to the original state i.e., before the photoexcitation. Now in the light of BGR we can explain this phenomenon as follows: the photoexcitation sets in BGR from the dense EHP and EHL formation within hundreds of fs. This features as excitonic bleach features at 610 nm and 670 nm in the initial timescale. In the case of BGR, the entire band structure is shifted to a lower energy and as a result, we see shifts in all the excitonic resonances as well. After the photoexcitation, the quasiparticles involved- excitons, EHL, EHP and phonons - generate intriguing dynamics involving the formation, inter-conversion and decay. These occur through various pathways including exciton formation from EHPs, excitons forming EHL, exciton evaporation from EHL surface, decay of excitons via single exciton recombination, exciton-exciton annihilation, single carrier recombination through nonradiative processes etc. The lower the pump fluence, the more the EHP population dominates over the EHL population and contribution of EHP increases in the total BGR. The EHP tend to form excitons within a few ps and the EHP population diminishes as well as the BGR from the EHP. Which we observe as the drastic shift in the A and B bleach features in the TAS spectra for the lower pump fluence of 0.26 mJ·cm⁻². Here, the renormalized band structure restores back to its original state to ~590 nm (B) and ~ 650 nm (A). But for the high fluence case this restoration of the bandgap is slower (few 10s of ps) due to the higher contribution of EHL in the population which have longer lifetimes (~ 20 ps). So, the restoration of the bandgap to its original band structure is slow or even unnoticeable in the higher pump fluences. We should mention in this context that the excitonic peaks in the fs-TAS spectra are somewhat shifted compared to the static absorption spectra where the excitonic resonance is at 610 nm (B) and 670 nm (A), whereas the excitonic resonances in the TAS spectra show at 590 nm (B) and 650 nm (A).

This is observed in various literature as well. In order to give an idea about the BGR, we compare TAS spectra for the pump fluences mentioned above, at fixed delays of 0.3 ps and 100 ps, respectively in panel **1c**. The red and blue dashed lines indicate the minima of the bleach signal from the B exciton at pump-probe delays of 0.3 ps and at 100 ps, respectively. The shift in the spectral features at the lower pump fluence is evident comparing these cases presented. Here, we observe that the line-shape of the TAS spectra is far from Gaussian or Lorentzian in shape. This is due to a number of factors, including a reduction in the oscillator strength, shift in excitonic resonance and broadening of line-width from various factors. The red-shift in excitonic resonance originates from the BGR arising out of increased carrier concentration following the pump excitation, as we depict in figure 1d. where EHPs and EHLs cause this BGR due to their increased charge carrier densities. In the forthcoming section we discuss the details of the analytical model used for understanding the contributions to the BGR at various time-scales as well as the contribution to the various species to the transient population as deduced from time-resolved spectra.

## 2. Theoretical model of the dynamics of excitons and excitonic condensates.

In figure 2, panel **a**, we present the amplitude $|\Delta A|$ of the bleach feature corresponding to the B exciton as a function of delay for various values of pump fluence from $0.26 - 10.2$ mJ·cm$^{-2}$ as indicated in the legend. To extract the power law dependence of the carrier population on the pump fluence, we plot $|\Delta A|$ as a function of the pump fluence for various time delays following the photoexcitation by the pump pulse and fit the corresponding data as a function of the fluence, $F$, with an exponent $\gamma$, which depends on probe delay as shown in panel b of figure 2. Evidently, for initial timescales, up to 1 ps the behavior of $\Delta A$ can be captured in the form of a power law $|\Delta A| \propto F^{\gamma}$. The fit yields $\gamma = 0.33$ for shorter delays while it decreases to $\gamma = 0.29$ for longer delays, cf. panel b. This power law behaviour motivates a model for the dynamics through rate equations governing the time-dependent populations in levels. The corresponding equation has a simple form: $dN/dt = -k_{rc}N - k_{ee}N^2$, where $N$ is the exciton density, $k_{ee}$ is the exciton-exciton annihilation rate and $k_{rc}$ is the exciton recombination rate. Here, the steady state solution of this equation yields an exponent of 0.5 in the power law dependence[13]. An exponent of less than 0.5 in the power law indicate a loss mechanism arising from faster decay channels. We also observe the following - as the amplitude reduces at a high pump fluence of 10.2 mJ·cm$^{-2}$, the fit does not quite match the experimental data at this regime. Since carrier trapping and other decay channels are a linear function of carrier concentration, we conclude that conversion to EHL-like multi-body aggregates is a possible path of free exciton (FE) loss resulting in the lower power exponent.

We now develop a model to predict the transient populations of different components - EHL, EHP, excitons and phonons- from the experimental data and discern their roles in the ensuing dynamics. In this model we employ a set of coupled rate equations to describe the interplay between the EHL, FE and the phonons; thus, we explain the condensation and evaporation of EHL droplets which are formed during the ultrafast relaxation of this photoexcited nanoscale system. We assume that as the exciton density exceeds a critical density, $N_c$, the condensation into the liquid-like phase is initiated i.e., beyond this critical carrier density the nucleation process begins. After the droplets are formed mainly two processes govern the dynamics of the droplet: **i)** the condensation of free excitons (FEs) on the droplet surface which leads to an increase in droplet volume and **ii)** the evaporation of FEs from the droplet surface which has the opposite effect of reducing droplet volume (shown in figure 3). Further, these carriers inside the droplet can also recombine, contributing to droplet shrinkage. Since we observed a decreasing power law behavior with an increase in pump-probe delay, we infer that the droplet population has a lifetime longer than that of FEs. This results in an abundance of EHL droplets around 10s of ps manifesting in a lower exponent for the measured |**ΔA**| with increasing pump fluence. Our system of interest comprises of few-layered MoS$_2$ NPs with a shell structure with

possibly inherent oxygen defect centers due to the laser ablation technique used to prepare them. These defect centers can trap the carriers and annihilate through non-radiative recombination and, thereby, we observe an increase in phonon population gradually as the trapped carriers annihilate[41] around 100s of ps specifically for higher pump fluences as depicted in panel **a** of figure 2.

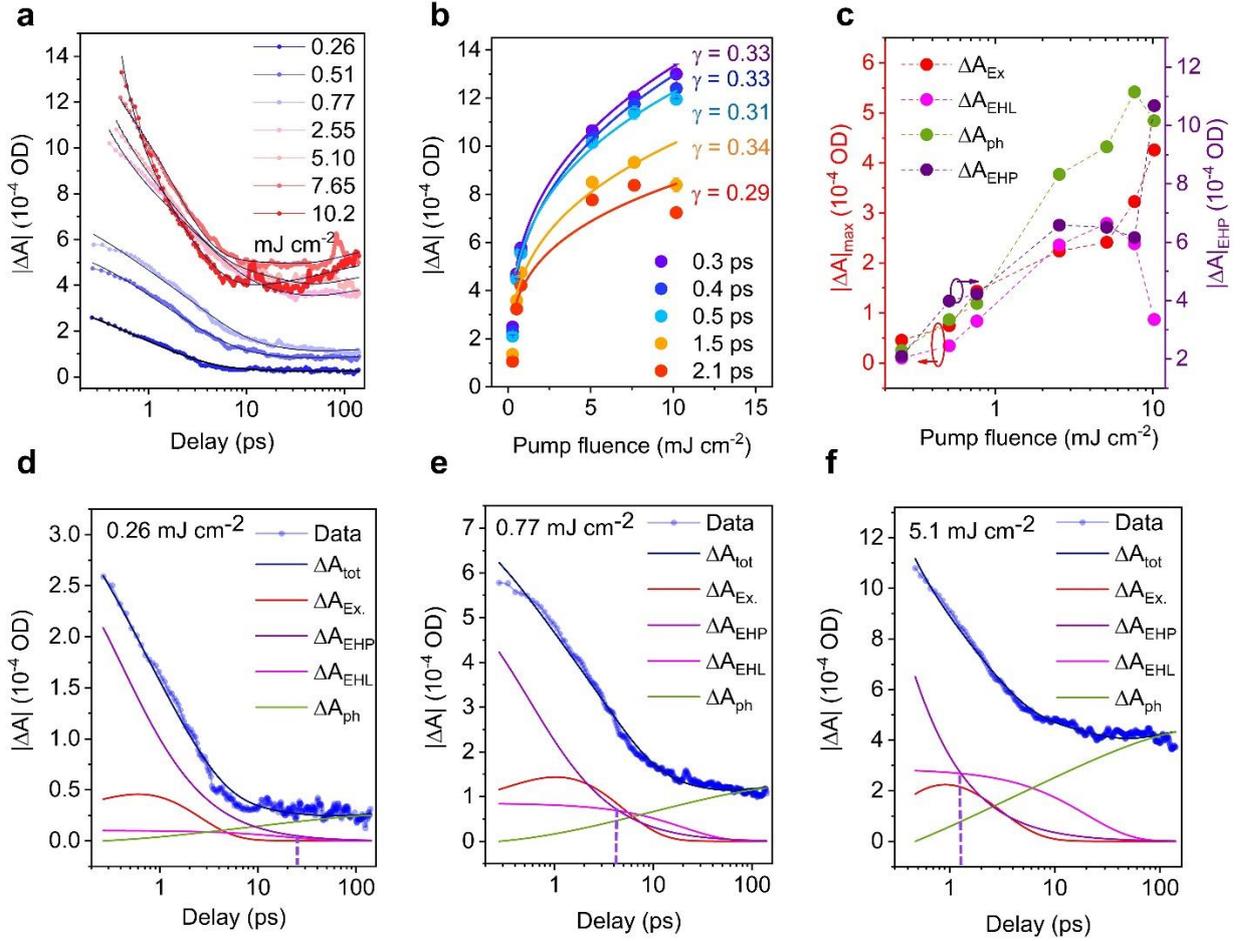

**Figure 2.** ((**Transient absorption spectra for exciton condensate dynamics**. **a**, Decay dynamics of the bleach feature corresponding to the B exciton at different pump fluences (solid circles), solid lines are a fit to experimental data using the proposed kinetic model (see text for details). **b**, Differential absorbance $|\Delta A|$ measured for the same feature as in panel a versus pump fluence for various time delays from 0.3 ps to 2.1 ps (solid circles). These are fit (solid lines) to the power-law detailed in the text with exponent $\gamma$. **c**, Maximum contributions of different excitonic species free excitons ($\Delta A_{Ex.}$), electron-hole plasma EHP ($\Delta A_{EHP}$), electron-hole liquid EHL ($\Delta A_{EHL}$) and phonons $\Delta A_{ph}$ in the total time-dependent absorbance $\Delta A$ as a function of pump fluence. These contributions to the total transient absorbance $|\Delta A|$ are deduced from numerical solutions of the coupled rate equations (1), (2) (3) and (4) for three different pump fluences as a function of delay 0.26 mJ·cm$^{-2}$, 0.77 mJ·cm$^{-2}$ and 5.1 mJ·cm$^{-2}$ in panels **d, e** and **f**, respectively. Solid dots are experimental data while in each case, the curves are derived from a fitting of the numerical solution to the proposed model. These discern the time-evolution of the contributors to the total differential absorption measured.))

In order to extract the respective populations of FEs, EHPs, EHLs and phonons, we use a dynamic model comprising the population densities for free excitons $N_{Ex.}(t)$, electron-hole pairs $N(t)$, the EHL $N_{EHL}(t)$ and phonons $N_{ph}(t)$ in the photoexcited volume of the sample containing MoS$_2$ nanoparticles in suspension. The relation and interdependence of these quantities is captured in the set of coupled differential equations[13,42]:

$$\frac{dN_{Ex}}{dt} = -\alpha N_{Ex.}^2 - k_2 N^2 - k_1 N_{Ex} - S_0(N_{Ex.}N_l) - \lambda_0(N_{Ex.} - N_c) \cdot \Theta(N_{Ex.} - N_c) \qquad (1)$$

$$\frac{dN_{EHL}}{dt} = \lambda_0(N_{Ex.} - N_c) \cdot \Theta(N_{Ex.} - N_c) + S_0 \cdot N_{Ex.}N_{EHL} - \frac{N_{EHL}}{\tau_{rec}} \qquad (2)$$

$$\frac{dN_{ph}}{dt} = k_{cap}(N + N_{EHL}) - \frac{N_{ph}}{\tau_{ph}} \qquad (3)$$

$$\frac{dN}{dt} = G(t) - k_2 N^2 - \frac{N}{\tau_{nr}} \qquad (4)$$

where, $\alpha$ is the exciton-exciton annihilation rate, $k_1$ is the rate of radiative recombination for excitons and $k_2$ is the electron-hole recombination rate. $k_1$ is extracted to be 0.25 - 0.58 ps$^{-1}$ while $k_2$ is between $0.09 \times 10^{-7}$ - $0.3 \times 10^{-7}$ cm$^3$s$^{-1}$ depending on the excitation fluence. $S_0 = 1 \times 10^{-16}$ cm$^3$s$^{-1}$ is the strength of the coupling between the exciton gas and the EHL droplet, $\lambda_0 = 1 \times 10^{-4}$ s$^{-1}$ is a constant factor describing the condensation of EHL as the exciton density $N_{\text{Ex.}}$ exceeds the critical density $N_c$, $\tau_{\text{rec}} = 20$ ps is the recombination lifetime of the carriers inside the EHL droplet[13], $k_{\text{cap}} = 0.1$ ps$^{-1}$ is the carrier capture rate in the defect sites, $\tau_{\text{ph}} = 10$ ns is the phonon decay lifetime, $\tau_{\text{nr}} = 100$ ps is the non-radiative recombination lifetime of the carriers. $G(t)$ is the carrier generation rate which is $G(t) = \frac{\eta I(t)}{\hbar \omega}$, where $I(t)$ is the incident pump excitation with a Gaussian profile centered at energy $\hbar \omega$; $\eta$ is the absorbance of the sample at the pump energy. However, after the passage of the pump pulse through the sample is complete, the generation term tends to zero i.e. $G(t) = 0$. We fit the data from 200 fs onwards where the effect of the pump pulse in generation of the carriers is nullified. Therefore, the coupled equations (1) - (4), $G(t) =$ capture all the essential features of the dynamics involving free excitons, EHP, EHL, phonons and free carriers after the initial free-carrier generation. Here, we consider the overall rate of two-body interactions in the model as $-\alpha N_{Ex}^2$ which include second-order excitonic interactions and formation of higher-order states such as bi-excitons, as well as decay rate via X-X annihilation. However, it is observed in literature that the bandgap renormalization/exciton binding energy (BGR/EBE) change dominates over the optically induced biexciton formation in monolayer MoS$_2$. As a result, we include the overall two-body interaction rate in our model which is dominated by X-X annihilation in our case. After solving the coupled rate equations, the differential absorbance $\Delta A$ is summed using the linear relation[42]:

$$\Delta A(t) = \alpha_{\text{Ex.}} N_{\text{Ex.}}(t) + \alpha_{\text{EHP}} N_{\text{EHP}}(t) + \alpha_{\text{EHL}} N_{\text{EHL}}(t) + \alpha_{\text{ph}} N_{\text{ph}}(t)$$

$$= \Delta A_{\text{Ex.}}(t) + \Delta A_{\text{EHP}}(t) + \Delta A_{\text{EHL}}(t) + \Delta A_{\text{ph}}(t) \tag{5}$$

The total time-dependent absorbance $\Delta A(t)$ thus determined is matched to experimental data and best-fit in an iterative process (equation 5). In this model, $\Delta A_{\text{Ex.}}(t)$ isolates the contribution from carrier-induced bleaching of the excitonic states while the contribution $\Delta A_{\mathbf{EHL}}(t)$ is the bleach signature from EHL, and $\Delta A_{\mathbf{ph}}(t)$ is the contribution from the defect-captured carriers which recombine generating phonons in the process. The contributions of these different components to the total $\Delta A$ are determined by fitting the measured TAS data with the proposed model. These contributions $\Delta A_{\mathbf{Ex.}}(t)$, $\Delta A_{\mathbf{EHP}}(t)$, $\Delta A_{\mathbf{EHL}}(t)$ and $\Delta A_{\mathbf{ph}}(t)$ are plotted in panel **2c** as a function of pump fluence. From this, it is obvious that as the fluence increases the fractional contribution to transient absorption from EHL increases by about 25 times while that of the free exciton increases by a factor of 9. Examining the components of contribution to composite $\Delta A(t)$ at three different pump fluences, i. e., 0.26, 0.78 and 5.2 mJ·cm⁻², respectively, depicted in figures **2d – f**, clearly evidence an increasing trend in the EHL-like phase along with the from trapped carriers with rising pump fluence. If we compare the time taken for the EHL contribution to exceed the EHP contribution, we see that for 0.26 mJ·cm⁻² it is ~11 ps whereas for 5.2 mJ·cm⁻² it is ~1 ps. This is indicated by a purple dashed line in panels 2d – f. It is clear that this time decreases as we increase the pump fluence hence, EHL starts to dominate the dynamics at earlier times for higher pump fluences.

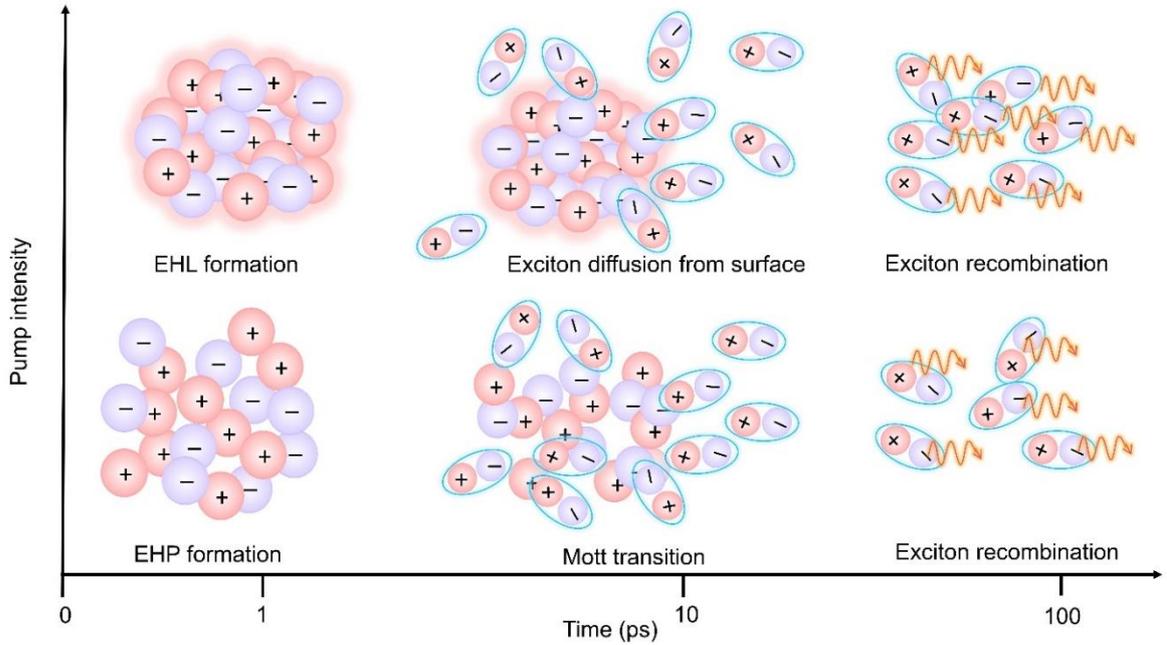

**Figure 3.** ((**2D schematic of carrier condensation and decay.** A depiction of the carrier condensation dynamics as a function of intensity along the vertical axis and temporal dynamics shown along horizontal axis. Electron-hole condensation and EHL droplet formation occurs with increasing pump intensity. Decay pathways of EHL and electron-hole plasma (EHP) are depicted with time. At lower pump fluences, only EHP and free excitons (FEs) are formed following photoexcitation followed by EHP transitioning into FE after a few picoseconds. Higher pump fluences generate carrier density above the critical density and carriers begin condensing into EHL droplets following photoexcitation. These EHL droplets gradually evaporate into FEs and possibly even recombine inside the droplet volume. Subsequently, FEs recombine through radiative and non-radiative means.))

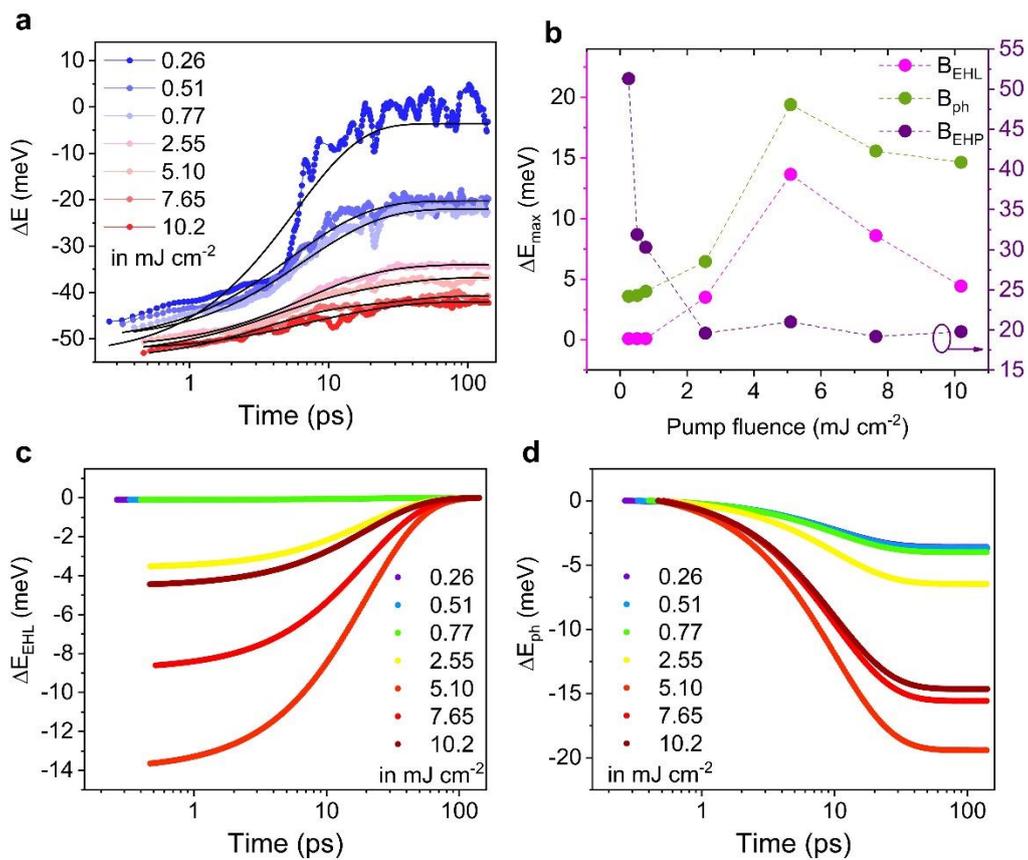

**Figure 4.** ((**Exciton resonance shifts and band gap renormalization. a,** Shift in the B excitonic resonance energy as a function of pump-probe delay at different pump fluences –

experimental data (solid circles) are fit (black solid lines) by solving the kinetic model for BGR as detailed in the text. **b**, Variations of the relative strengths of BGR corresponding to EHP (violet circles), EHL (pink circles) and phonon (green circles) as a function of pump fluence. Wbile EHL and phonon strengths correspond to the left vertical axis and EHP BGR strength is referred to the right. **c**, Temporal evolution of the EHL contribution to the BGR, $\Delta E_{\text{EHL}}(t)$, andd the temporal evolution of the phonon contribution. Panels **c** and **d** establish that the dominant contribution to the BGR on timescale of 1-10 ps is due to the formation of the EHL-phase ahead of phonons. On much longer timescales, $10 - 100$ ps, this diminishes and gives way to phonons, eventually, which take over the dynamics.))

**Band-gap renormalization from EHP, EHL and phonons**

All these excitonic condensed phases will induce and contribute to the screening of the Coulomb interaction between the constituent charges, resulting in an effective bandgap renormalization (BGR). This is reflected in our TAS measurements as a red shift of the excitonic resonance energies at early timescales as depicted in figure 4a. The variations of the relative strengths of BGR corresponding to EHP , EHL and phonons as a function of pump fluence is shown in figure 4b. Although BGR due to EHP[37,43,44] and phonons[42] is well explored in different systems, that arising out of the screening from EHL remains largely uninvestigated. From time-dependent absorption measurements of the total differential absorption $\Delta A(t)$ and our physical model, clearly, a liquid-like phase in the excitonic aggregate along with other contributing factors is important and remain as irreplaceable factor in accounting for the observed dynamics. Correspondingly in the case of time-dependent BGR shifts we observe that the EHL phase contributes dominantly to the observed shifts of on timescales of 1-10 ps. A quick comparison of panels 4c and 4d, where the respective BGR contributions of the EHL and phonons are plotted as a function of time leads us to this emphatic conclusion: We model the total observed BGR to be a cumulative effect of EHP, EHL and phonon contributions, $\Delta E = \Delta E_{EHP} + \Delta E_{EHL} + \Delta E_{ph}$. Here, the contribution of the EHP component is calculated by solving the nonlinear differential equation[37]

$$\frac{d(\Delta E_{EHP})}{dt} = B_{EHP} k_1 e^{-k_1 t} - \frac{k_2 N_0}{B_{EHP}} \Delta E_{EHP}^2 \qquad (6)$$

while assuming the other contributions to follow exponential relations[37]

$$\Delta E_{EHL} = B_{EHL} e^{-k_{rec} t} \qquad (7)$$

and

$$\Delta E_{ph} = B_{ph}\left(1 - e^{-k_{ph} t}\right) \qquad (8)$$

where, $B_{EHP}$, $B_{EHL}$ and $B_{ph}$ are the fitting parameters which capture the strengths of relative contributions of the EHP, EHL and phonon contributions to the BGR. The values of the rate constants $k_1$, $k_2$, $k_{rec}$ and $k_{ph}$ are the same as those obtained by solving equations 1) – 4) earlier; $N_0$ is the initial carrier concentration. Using equations 6) – 8), we discern the contributions of the 3 key phases of the excitonic aggregate which leads to the following picture emerging from the BGR dynamics: The pump pulse creates a dense EHP transition directly to free excitons at carrier densities *below* the Mott density at 0.26 mJ·cm⁻². This is evidenced by a step-like rise in the BGR around 4 ps. However, as the fluence is raised all the way to 10 mJ·cm⁻² (about 40 times the pump fluence of 0.26 mJ·cm⁻²) this gives way to a more gradual change in the BGR clearly evidencing the formation of a new phase between the dense plasma and the dilute free gas. This is the liquid-like phase which plays a crucial role in the dynamics on the few ps timescale. On these timescales we observe the rise in the contribution of the differential absorption arising out of this novel phase to the total observed differential transient absorption, cf. panels **d** – **f** in figure 2. Therefore, the EHL phase makes its dominant contribution to the BGR, as it does in the case of the differential transient absorption, until ~10 ps of relaxation following the photoexcitation by the pump pulse. Thereafter, the time-dependent absorption measurements suggest that the liquid phase decays and tails off together with EHP. Phononic contributions to the BGR kick-in at this stage and take over the dynamics on 10 – 100 ps timescales. This completes the life cycle of the excitonic aggregate in this unique system where we evidence the formation, dynamics and decay of the liquid-like EHL phase. This insight into the important features of ultrafast dynamics is very important in designing devices such as high-power, high-frequency lasers and LEDs.

## 3. Conclusion

In summary, we demonstrate the formation of a quantum liquid-like EHL phase at high pump-fluences in this TMDC nanoparticle systems at room temperature. The interplay between FE, EHL and EHP is modeled and found to fit the experimental TAS data quite well allowing us to discern the existence of the novel liquid phase. After intense photoexcitation, the band structure of TMDCs undergoes an extensive renormalization from the excitonic many-body bound states such as electron-hole liquid along with ionized electron-hole plasma. A BGR of about 50 meV is observed for the highest pump fluence. This BGR is a result of the strong Coulomb interaction between the carriers at such high intensities. The dynamics of the EHL and EHP phases are analyzed from the temporal evolution of excitonic resonances. Explorations over a broad range of pump powers allows us to obtain insights into the individual contributions of the droplet, EHP, excitonic and phononic states in the total transient absorbance as well the BGR. From this, the following generic picture emerges: the dynamics is dominated by the EHP phase at initial timescales of 2-4 ps, after which the quantum EHL phase emerges and starts to dominate the transient absorption till ~ 10 ps. Within this period, the trapping of carriers results in phonon emission, which subsequently takes over the dynamics until these carriers decay via radiative or non-radiative processes. These TAS measurements, over an ensemble of NPs, enable accurate spectral sampling even at low-excitation powers. The room-temperature realization of the quantum EHL state renders TMDC nanoparticle systems as an ideal candidate for exploring many-body interactions and macroscopic quantum states under ambient conditions, at a fundamental level. It is interesting to note that the quantum nature of EHLs in inter-layer indirect excitonic clusters which has been demonstrated in a recent article[6], further emphasizes its applicability in quantum circuits. The results of our numerical modeling and experiments on the formation and evolution of the excitonic condensates lead to identification of this new paradigm, opening up ample scope for harnessing the electron-hole liquid at room-temperature, in likely quantum devices.

## 4.    Methods

**4.1.    MoS$_2$ nanoparticle sample preparation.** MoS$_2$ nanoparticles were prepared using a top-down method - femtosecond pulsed laser ablation in liquids (fs-PLAL). The first step of this process involved preparing MoS$_2$ pellets were prepared from MoS$_2$ powder (Sigma Aldrich) using hot press sintering at 100° C by applying a pressure of 5 ton. The pellet was ablated in 30 ml of deionized (DI) water using 1 ps pulses of 1.5 W average power, generated by a Ti:Sapphire femtosecond laser system (Astrella, Coherent Inc.). The ablation time, wavelength and repetition rate were kept constant at 20 min, 800 nm, and 1 kHz, respectively. The ablated suspension was centrifuged at 2500 rpm for 15 minutes. The supernatant (top 1/3 portion of the centrifuged solution) was extracted for further experiments. To confirm the sample quality, steady-state absorption, photoluminescence (PL), HRTEM and Raman spectroscopy were performed, as detailed in the supplementary section 1.

**4.2.    Femtosecond transient absorption spectroscopy (fs-TAS).** Transient absorption data is obtained by exciting the sample, suspended in deionized water, with a non-resonant 114 fs, 500 nm (2.48 eV) pump pulse and white light probe generated from a Sapphire plate with a wavelength span of 500-800 nm, using the HELIOS transient absorption spectrometer (Ultrafast systems, USA). Pump fluences used in this study ranges from 0.26 to 10.2 mJ·cm$^{-2}$. Details of the carrier density calculations and the TAS data analysis method is described in supplementary sections 2, 3 and 4.

### 4.3. Computational methods.

We have implemented all theoretical calculations within the framework of plane-wave density functional theory (PW-DFT) using the Quantum Espresso code[45]. To match obtained experimental bandgap, we used the generalized gradient approximation along with Hubbard parameter (GGA+U: U=4.5 eV) in all calculations[46]. The optical properties with their absorption coefficients of $MoS_2$ were obtained through complex dielectric functions using the formalism of Ehrenreich and Cohen[47]. From these, the optical absorption coefficient $\alpha(\omega)$ was computed as

$$\alpha(\omega) = \sqrt{2}\omega \left( \sqrt{\varepsilon_1(\omega)^2 + \varepsilon_2(\omega)^2} - \varepsilon_1(\omega) \right)^{1/2} \tag{9}$$

The details of the computational method are presented in the supplementary section 2.

ASSOCIATED CONTENT

**Associated content**

Supplementary document.docx contains details about the synthesis and characterization of nanoparticles, density functional theory calculations and estimation of photo-induced carrier density, details of femtosecond transient absorption spectroscopy measurements as well as that of the estimation of electron-hole droplet (EHD) size and number per nanoparticle.

**Acknowledgements**

SRK acknowledges financial support from the Science and Engineering Research Board, Department of Science and Technology, Govt. of India, from the Institute of Eminence scheme through the Quantum Center for Diamond and Emergent Materials and Micro-Nano-Bio-Fluidics prospective centers of excellence, from the Max Planck Society, Germany through the Partner Group programme, from the Scheme for Promotion of Academic and Research

Collaboration, Ministry of Education, India. PD acknowledges financial support from the Indian Institute of technology Madras through the Women Leading IITM (WLI) scheme. All authors acknowledge the support of the Femto Science Facility, Dept. of Physics, IIT-Madras, for providing unrestricted access to both the fs laser and TAS systems.

**Author contributions**

The manuscript was written through contributions of all authors. All authors have given approval to the final version of the manuscript. P.D. contributed to performing the transient absorption measurements, parts of spectroscopic and microscopic characterizations, data analysis and manuscript writing. T.D. contributed to data analysis and interpretation, as well as manuscript preparation. V. M. did the DFT calculations and contributed to manuscript preparation. A. S. prepared the nanoparticle samples for by fs pulse ablation and performed spectroscopic characterization measurements, was instrumental in handling the fs laser and optical parametric amplifier. C. V. contributed to manuscript preparation and interpretation of results. S. R. K. conceived the project plan, led this effort as the principal investigator and was responsible for generating resources for research, contributed to experiments, data analysis and interpretation of results and manuscript preparation.

**Competing interests**

The authors declare no competing interests.

**Supporting Information**

Supporting Information is available from the Wiley Online Library or from the author.